\documentclass[10pt]{article}

\usepackage{amsxtra,amssymb,amsthm,amsmath,latexsym}

\usepackage{graphicx}

\newcommand{\R}{{\mathbb R}}

\newcommand{\bee}{\begin{equation*}}
\newcommand{\eee}{\end{equation*}}
\newcommand{\be}{\begin{equation}}
\newcommand{\ee}{\end{equation}}
\newcommand{\pn}{\par\noindent}
\title{Scattering of electromagnetic waves by many nanowires
 }
\author{A G Ramm\\
\small Department of Mathematics\\[-0.8ex]
\small Kansas State University, Manhattan, KS 66506-2602, USA\\[-0.8ex]
\small \texttt{ramm@math.ksu.edu}\\
}
\date{}
\begin{document}

\maketitle
\begin{abstract}
Electromagnetic wave scattering by many parallel to $z-$axis, thin,
impedance, circular infinite cylinders
is studied asymptotically as $a\to 0$.
Let  $D_m$ be the crossection
of the $m-$th cylinder, $a$ be its radius, and
$\hat{x}_m=(x_{m1},x_{m2})$ be its
center, $1\le m \le M$, $M=M(a)$.
It is assumed that the points $\hat{x}_m$ are distributed
so that
$$\mathcal{N}(\Delta)=\frac 1{2\pi
a}\int_{\Delta}N(\hat{x})d\hat{x}[1+o(1)],
$$
where $\mathcal{N}(\Delta)$ is the number of points
$\hat{x}_m$ in an arbitrary open subset $\Delta$ of the plane
$xoy$. The function $N(\hat{x})\geq 0$ is
a given continuous function. An equation for the self-consistent
(efficient) field is derived as $a\to 0$.
A formula is derived for the effective refraction
coefficient in
the medium in which many thin impedance cylinders are distributed. These
cylinders may
model nanowires embedded in the medium. Our result shows how these
cylinders influence the refraction coefficient of the medium.
\end{abstract}
\pn{
 }

\section{Introduction}
There is a large literature on electromagnetic (EM) wave scattering
by an array of parallel cylinders (see, e.g., \cite{M}, where there
are many references given, and \cite{MV}).
Electromagnetic wave scattering by many parallel to $z-$axis, thin,
circular, of radius $a$, infinite cylinders,  on the boundary of which
an impedance boundary condition holds,
is studied in this paper asymptotically as $a\to 0$.
 The cylinders are thin in the sense $ka\ll 1$,
where $k$ is the wave number in the exterior of the cylinders,

The novel points in this paper include:

1) The solution to the wave scattering problem is considered  in
the limit $a\to 0$ when the number $M=M(a)$ of the cylinders tends
to infinity at a suitable rate. The equation for the limiting (as
$a\to 0$) effective (self-consistent) field in the medium is
derived,

2) This theory is a basis for a method for changing
refraction coefficient in a medium. The thin cylinders model nanowires
embedded in the medium. The basic physical resul of this paper is
formula  \eqref{e47}, which shows how the embedded thin cylinders change
the refraction coefficient $n^2(x)$.

Some extension of the author's
results (\cite{R515}-\cite{R595}) is obtained for EM wave
scattering
by many thin perfectly conducting cylinders. The techniques used are
similar to the ones developed in \cite{R610}.

Let $D_m$, $1\leq m\leq M$, be a set of non-intersecting domains on
a plane $P$, which is $xoy$ plane. Let $\hat{x}_m\in D_m$,
$\hat{x}_m=(x_{m1},x_{m2})$, be a point inside $D_m$ and $C_m$ be
the cylinder with the cross-section $D_m$ and the axis, parallel to
$z$-axis, passing through $\hat{x}_m$. We assume that $\hat{x}_m$
is the center of the disc $D_m$ if $D_m$ is a disc of radius $a$.

Let us assume  that on the boundary of the cylinders an imedpance
boundary condition holds, see \eqref{e5} below. Let
$a=0.5$diam$D_m$. Our basic assumptions are
\be\label{e1} ka\ll 1,
\ee where $k$ is the wave number in the region exterior to the union
of the cylinders, and \be\label{e2} \mathcal{N}(\Delta)=
\frac{1}{2\pi a}\int_{\Delta}N(\hat{x})d\hat{x}[1+o(1)],\quad a\to 0, \ee
where $\mathcal{N}(\Delta)=\sum_{\hat{x}_m\in \Delta}1$ is the
number of the cylinders in an arbitrary open subset of the plane
$P$, $N(\hat{x})\geq 0$ is a continuous function, which can be
chosen as we wish, and $2\pi a$ is the arclength of a circle of radius
$a$. The points $\hat{x}_m$ are distributed in an arbitrary
large but fixed bounded domain on the plane $P$. We denote by
$\Omega$ the union of domains $D_m$, by $\Omega'$ its complement in
$P$, and by $D'$ the complement of $D$ in $P$. The complement in
$\R^3$ of the union $C$ of the cylinders $C_m$ we denote by $C'$.

The EM wave scattering problem consists of finding the solution to
Maxwell's equations

\be\label{e3} \nabla \times E=i\omega \mu H, \ee \be\label{e4}
\nabla \times H=-i\omega \epsilon E, \ee in $C'$, such that
\be\label{e5} E_t=\zeta[n,H] \text{ on } \partial C, \ee where
$\partial C$ is the union of the surfaces of the cylinders $C_m$,
$E_t$ is the tangential component of $E$ on the boundary of $C$, $n$
is the unit normal to $\partial C$ directed out of the cylinders,
$\mu$ and $\epsilon$ are constants in $C'$, $\omega$ is the
frequency, $k^2=\omega^2\epsilon \mu$, $k$ is the wave number.
Denote by $n_0^2=\epsilon \mu$, so $k^2=\omega^2 n_0^2$. The
solution to \eqref{e3}-\eqref{e5} must have the following form
\be\label{e6} E(x)=E_0(x)+v(x),\quad
x=(x_1,x_2,x_3)=(x,y,z)=(\hat{x},z), \ee where $E_0(x)$ is the
incident field, and $v$ is the scattered field satisfying the
radiation condition \be\label{e7} \sqrt{r}\left(\frac{\partial
v}{\partial r}-ikv\right)=o(1),\quad r=(x_1^2+x_2^2)^{1/2}, \ee and
we assume that \be\label{e8} E_0(x)=k^{-1}e^{i\kappa y+i
k_3z}(-k_3e_2+\kappa e_3),\quad \kappa^2+k_3^2=k^2, \ee $\{e_j\}$,
$j=1,2,3$, are the unit vectors along the Cartesian coordinate axes
$x,y,z$. We consider EM waves with $H_3:=H_z=0$, i.e., E-waves, or
TH waves, \be\label{e9} E=\sum_{j=1}^3E_je_j,\quad
H=H_1e_1+H_2e_2=\frac{\nabla\times E}{i\omega \mu}. \ee One can
prove (see Appendix)  that the components of $E$ can be expressed by
the formulas: \be\label{e10}
E_j=\frac{ik_3}{\kappa^2}U_{x_j}e^{ik_3z},\quad j=1,2,\quad
E_3=Ue^{ik_3z},\quad U=\frac {\kappa}{k}u,  \ee where
$u_{x_j}:=\frac{\partial u}{\partial x_j}$, $u=u(x,y)$ solves the
problem \be\label{e11} (\Delta_2+\kappa^2)u=0\text{ in }\Omega' \ee
\be\label{e12} (u_n+i\xi u)|_{\partial \Omega}=0, \quad u_n:=\nabla
u\cdot n,\quad \xi:=\frac{\omega \mu \kappa^2}{\zeta k^2} \ee
\be\label{e13} u=e^{i\kappa y}+w, \ee and $w$ satisfies the
radiation condition \eqref{e7}. 


The unique solution to
\eqref{e3}-\eqref{e8} is given by the formulas: \be\label{e14}
E_1=\frac{ik_3}{\kappa^2}U_xe^{ik_3z} ,\quad
E_2=\frac{ik_3}{\kappa^2}U_ye^{ik_3z},\quad E_3=Ue^{ik_3z},\ee
\be\label{e15} H_1=\frac{k^2}{i\omega \mu
\kappa^2}U_ye^{ik_3z},\quad H_2=-\frac{k^2}{i\omega \mu \kappa^2}U_x
e^{ik_3 z},\quad H_3=0, \ee where $U_x:=\frac{\partial U}{\partial
x}$, $U_y$ is defined similarly, and $u=u(\hat{x})=u(x,y)$ solves
scalar two-dimensional problem \eqref{e11}-\eqref{e13}.  These
formulas are derived in the Appendix for convenience of the reader.

Problem \eqref{e11}-\eqref{e13} has
a unique solution (see, e.g., \cite{R190}) provided that Re$\zeta \ge 0$,
or, equivalently, that Im$\xi \ge 0$.
This corresponds to the assumption that the material inside the cylinders
is passive, that is, the energy absorption is non-negative.
Our goal is to derive an asymptotic
formula for this solution as $a\to 0$. Our results
include formulas for the solution to the scattering problem,
derivation of the equation for the effective field in the medium
obtained by embedding many thin perfectly conducting cylinders,
and a formula for the refraction coefficient in this limiting medium.
This formula shows that by choosing suitable distribution of the
cylinders, one can change the refraction coefficient, one can make
it smaller than the original one.

The paper is organized as follows.

In Section 2 we derive an
asymptotic formula for the solution to \eqref{e11}-\eqref{e13} when
$M=1$, i.e., for scattering by one cylinder.

In Section 3 we derive
a linear algebraic system for finding some numbers that define the
solution to problem \eqref{e11}-\eqref{e13} with $M>1$. Also in
Section 3 we derive an integral equation for the effective
(self-consistent) field in the medium with $M(a)\to \infty$
cylinders as $a\to 0$. At the end of Section 3 these results are
applied to the problem of changing the
refraction coefficient of a given material by embedding many
thin perfectly conducting cylinders into it.

In Section 4 conclusions are formulated.

In Appendix formulas \eqref{e14}-\eqref{e15} are derived.

\section{EM wave scattering by one thin perfectly conducting cylinder}
Consider problem \eqref{e11}-\eqref{e13} with $\Omega=D_1$,
$\Omega'$ being the complement to $D_1$ in $\R^2$. Assume for simplicity
that
$D_1$ is a circle $x_1^2+x_2^2\leq a^2$.

Let us look for a solution of the form \be\label{e16} u=e^{i\kappa
y}+\int_{S_1}g(\hat{x},t)\sigma(t)dt,\qquad
g(\hat{x},t):=\frac{i}{4}H_0^{(1)}(\kappa|\hat{x}-t|), \ee where
$S_1$ is the boundary of $D_1$, $H_0^{(1)}$ is the Hankel function
of order $1$, with index $0$,  and $\sigma$ is to be found from the
boundary condition \eqref{e12}. It is known that as $r\to 0$ one has
\be\label{e17} g(\kappa r)=\alpha(\kappa)+\frac{1}{2\pi}\ln
\frac{1}{r}+o(1),\quad
\alpha:=\alpha(\kappa):=\frac{i}{4}+\frac{1}{2\pi}\ln
\frac{2}{\kappa}, \ee and \be\label{e18} g(kr)=\frac i 4 \sqrt{\frac
2{\pi k r}}e^{i(kr-\frac {\pi}{4})} \big(1+o(1)\big), \quad r\to
\infty.
 \ee
Thus,
\be\label{e17'}
u=u_0+g(\hat{x},0)Q+o(\frac 1 {\sqrt{r}}),\quad r\to \infty; \quad
Q:=\int_{S_1}\sigma(t)dt.
\ee
The condition $r\to \infty$ is satisfied numerically if $r>>a$.
Consequently, it is sufficient to find one number $Q$ in order so solve
the scattering problem \eqref{e11}-\eqref{e13} for one thin cylinder.
The function \eqref{e16} satisfies equations
\eqref{e11} and \eqref{e13} for any $\sigma$, and if $\sigma$ is
such that function \eqref{e16} satisfies boundary condition
\eqref{e12}, then $u$ solves problem \eqref{e11}-\eqref{e13}.
We assume $\sigma$ sufficiently smooth (H\"older-continuous is
sufficient).

The solution to problem \eqref{e11}-\eqref{e13} is known to be
unique (see, e.g., \cite{R190}). Boundary condition \eqref{e12}
yields \be\label{e19} -u_{0n}(s)-i\xi u_0=i\xi \alpha Q +i\xi
\int_{S_1}g_0(s,t)\sigma(t) dt+ (A\sigma-\sigma)/2 , \ee where
$A\sigma:=\int_{S_1}\frac {\partial g_0(s,t)}{\partial n_s}dt,$ the
formula for the limiting value on $S_1$ of the exterior normal
derivative of the simple layer potential
$\int_{S_1}g_0(x,t)\sigma(t) dt$ was used, and
\be\label{e20}
u_0(s):=e^{i\kappa s_2}, \ s\in S_1;\,\,\,
g_0(s,t)=\frac{1}{2\pi}\ln\frac{1}{r_{st}},\,\, r_{st}:=|s-t|.\ee If
$ka\ll 1$, $k^2=\kappa^2+k_3^2$, then
 \be\label{e20'}
u_0(s)=1+O(\kappa a), \quad u_{0n}=i\kappa n_2+O(\kappa a). \ee

Equation \eqref{e19} is uniquely solvable for $\sigma$ if $a$ is
sufficiently small (see \cite{R476}).

We are interested in finding
asymptotic formula for $Q$ as $a\to 0$, because $u(\hat{x})$ in
\eqref{e16}
can be well approximated in the region $|\hat{x}|\gg a$ by the
formula \be\label{e21}
u(\hat{x})=u_0(\hat{x})+g(\hat{x},0)Q+o(1),\quad a\to 0.
\ee To find asymptotic of $Q$ as $a\to 0$, let us integrate equation
\eqref{e19} over $S_1$, keep the main terms of the asymptotic as
$a\to 0$, take into account that
$$\int_{S_1}dtN_2(t)=0,\quad
\int_{S_1}g_0(s,t)ds=O(a|\log a|) \qquad a\to 0,$$ use formulas  
\eqref{e20'}, and
obtain \be\label{e22} Q=i\xi u_0(\hat{x}_1) |S_1|, \ee where
$\hat{x}_1$ is a point inside $D_1$, $|S_1|$ is the length of $S_1$,
$|S_1|=2\pi a$ if $S_1$ is the circle $|\hat{x}|=a$, and $
r_{st}=|s-t|$. The reader can find the proof of the estimate
$\int_{S_1}g_0(s,t)ds=O(a|\log a|)$ as $a\to 0$, where $s,t\in S_1$, in
\cite{R610}. From formulas \eqref{e22} and \eqref{e17'} the
asymptotic solution to the scattering problem
\eqref{e11}-\eqref{e13} in the case of one circular cylinder of
radius $a$, as $a\to 0$, is \be\label{e28} u(\hat{x})\sim
u_0(\hat{x})+i2\pi a \xi g(\hat{x},0)u_0(\hat{x}_1),\,\,\, a\to
0,\,\,\, |\hat{x}|>a. \ee Electromagnetic wave, scattered by the
single cylinder, is calculated by formulas \eqref{e14}-\eqref{e15}
in which $u=u(\hat{x}):=u(x_1,x_2)$ is given by formula \eqref{e28}.

\section{Wave scattering by many thin cylinders}
Problem \eqref{e11}-\eqref{e13} should be solved  when
$\Omega$ is a union of many small domains $D_m$,
$\Omega=\cup_{m=1}^M D_m$. We assume  that $D_m$
is a circle of radius $a$ centered at the point $\hat{x}_m$.

Let us
look for $u$ of the form \be\label{e29}
u(\hat{x})=u_0(\hat{x})+\sum_{m=1}^M\int_{S_m}g(\hat{x},t)\sigma_m(t)dt.
\ee We assume that the points $\hat{x}_m$ are distributed in a
bounded domain $D$ on the plane $P=xoy$ by formula \eqref{e2}. The
 field $u_0(\hat{x})$ is the same as in Section 2,
$u_0(\hat{x})=e^{i\kappa y}$,
and Green's function  $g$ is the same as in
formulas
\eqref{e16}-\eqref{e18}. It follows from \eqref{e2} that
$M=M(a)=O(\ln \frac{1}{a}).$ We define the effective field, acting
on the $D_j$ by the formula \be\label{e30}
u_e=u_e^{(j)}=u(\hat{x})-\int_{S_j}g(\hat{x},t)\sigma_j(t)dt,\quad
|\hat{x}-\hat{x}_j|>a,
\ee
which can also be written as
$$u_e(\hat{x})=u_0(\hat{x})+
\sum_{m=1, m\neq j}^M\int_{S_m}g(\hat{x},t)\sigma_m(t)dt.$$
We assume that the distance $d=d(a)$
between neighboring cylinders   is much greater
than $a$: \be\label{e31} d\gg a, \quad \lim_{a\to
0}\frac{a}{d(a)}=0. \ee

Let us rewrite \eqref{e29} as \be\label{e32} u=u_0+\sum_{m=1}^M
g(\hat{x},\hat{x}_m)Q_m+\sum_{m=1}^M\int_{S_m}[g(\hat{x},t)-g(\hat{x},\hat{x}_m)]\sigma_m(t)dt,
\ee where \be\label{e33} Q_m:=\int_{S_m}\sigma_m(t)dt. \ee As $a\to
0$, the second sum in \eqref{e32} (let us denote it $\Sigma_2$)
is negligible compared
with the first sum in \eqref{e32}, denoted $\Sigma_1$,
\be\label{e34}
|\Sigma_2|\ll
|\Sigma_1|,\quad a\to 0. \ee
The idea of the  proof of this is similar to the one given in \cite{R509}
for a problem in $\R^3$. let us sketch this proof.

Let us check that \be\label{e35}
|g(\hat{x},\hat{x}_m)Q_m|\gg
|\int_{S_m}[g(\hat{x},t)-g(\hat{x},x_m)]\sigma_m(t)dt|,\quad a\to 0.
\ee If $k|\hat{x}-\hat{x}_m|\gg 1$, and $k>0$ is fixed then
$$|g(\hat{x},\hat{x}_m)|=O(\frac{1}{|\hat{x}-\hat{x}_m|^{1/2}}), \qquad
|g(\hat{x},t)-g(\hat{x},x_m)|=O(\frac{a}{|\hat{x}-\hat{x}_m|^{1/2}}),$$
and $Q_m\neq 0$, so estimate \eqref{e35} holds.

If
$$|\hat{x}-\hat{x}_m|\sim
d\gg a,$$ then
$$|g(\hat{x},\hat{x}_m)|=O(\frac{1}{ln \frac{1}{a}}), \qquad
|g(\hat{x},t)-g(\hat{x},x_m)|=O(\frac{a}{d}),$$ as follows from the
asymptotics of $H_0^1(r)=O(\ln \frac{1}{r})$ as $r\to 0$, and
from the formulas $\frac{dH_0^1(r)}{dr}=-H_1^1(r)=O(\frac{1}{r})$ as $r\to
0$.
Thus,
\eqref{e35} holds for $|\hat{x}-\hat{x}_m|\gg d\gg a$.

Consequently, the scattering problem is reduced to finding the
numbers $Q_m$, $1\leq m\leq M$.

Let us estimate $Q_m$ asymptotically, as $a\to 0$. To do this, we
use the exact boundary condition on $S_m$ and an arguments similar
to the one given in the case of wave scattering by one cylinder. The
role of the incident field $u_0$ is played now by the effective
field $u_e$. The result is a formula, similar to \eqref{e22}:
\be\label{e38} Q_j=i2\pi a \xi u_e(\hat{x}_j), \quad a\to 0. \ee
Formula, similar to \eqref{e28}, is \be\label{e39} u(\hat{x})\sim
u_0(\hat{x})+i2\pi a \xi
\sum_{m=1}^Mg(\hat{x},\hat{x}_m)u_e(\hat{x}_m), \quad a\to 0. \ee
The numbers $u_e(\hat{x}_m)$, $1\leq m\leq M$, in \eqref{e39} are
not known. Setting $\hat{x}=\hat{x}_j$ in \eqref{e39}, neglecting
$o(1)$ term, and using the definition \eqref{e30} of the effective
field, one gets a linear algebraic system for finding numbers
$u_e(\hat{x}_m)$: \be\label{e40} u_e(\hat{x}_j)=u_0(\hat{x}_j)+i2\pi
a \xi \sum_{m\neq j}g(\hat{x}_j,\hat{x}_m)u_e(\hat{x}_m),\quad 1\le
j\le M. \ee This system can be solved numerically if the number $M$
is not very large, say $M\leq O(10^3)$.

If $M$ is very large, $M=M(a)\to \infty,\,\, a\to 0$,  then we derive
a linear
integral equation for the limiting effective field in the medium
obtained by embedding many cylinders.

Passing to the limit $a\to 0$ in system \eqref{e40} is done as in
\cite{R610}. Consider a partition of the domain $D$ into a union of
$\bf{P}$ small squares $\Delta_p$, of size $b=b(a)$, $b\gg d\gg a$.
For example, one may choose $b=O(a^{1/4})$, $d=O(a^{1/2})$, so that
there are many discs $D_m$ in the square  $\Delta_p$. We assume that
squares $\Delta_p$ and $\Delta_q$ do not have common interior points
if $p\neq q$. Let $\hat{y}_p$ be the center of $\Delta_p$. One can
also choose as $\hat{y}_p$  any point $\hat{x}_m$ in a domain
$D_m\subset \Delta_p$. Since $u_e$ is a continuous function, one may
approximate $u_e(\hat{x}_m)$ by $u_e(\hat{y}_p)$, provided that
$\hat{x}_m\subset \Delta_p$. The error of this approximation is
$o(1)$ as $a\to 0$. Let us rewrite the sum in \eqref{e40} as
follows: \be\label{e41} 2\pi a\sum_{m\neq
j}g(\hat{x}_j,\hat{x}_m)u_e(\hat{x}_m)=
\sum^{\bf{P}}_{\stackrel{p=1}{x_j\notin
\Delta_p}}g(\hat{x}_j,\hat{y}_p)u_e(\hat{y}_p) 2\pi a \sum_{x_m\in
\Delta_p}1,\ee and use formula \eqref{e2} in the form \be\label{e42}
2\pi a \sum_{x_m\in \Delta_p}1=N(\hat{y}_p)|\Delta_p|[1+o(1)], \quad
a\to 0. \ee Here $|\Delta_p|$ is the volume of the square
$\Delta_p$.

From \eqref{e41} and \eqref{e42}
one obtains: \be\label{e43} 2\pi a \sum_{m\neq
j}g(\hat{x}_j,\hat{x}_m)u_e(\hat{x}_m)=
 \sum^{\bf{P}}_{\stackrel{p=1}{\hat{x}_j\notin
\Delta_p}}
g(\hat{x}_j,\hat{y}_p)N(\hat{y}_p)u_e(\hat{y}_p)|\Delta_p|[1+o(1)].\ee
The sum in the right-hand side of formula \eqref{e43} is the
Riemannian sum for the integral \be\label{e44} \lim_{a\to
0}\sum_{p=1}^{\bf{P}}g(\hat{x}_j,\hat{y}_p)N(\hat{y}_p)u_e(\hat{y}_p)|\Delta_p|=\int_D
g(\hat{x},\hat{y})N(\hat{y})u(\hat{y})dy,\quad u(\hat{x})=\lim_{a\to
0}u_e(\hat{x}). \ee Therefore, system \eqref{e40} in the limit $a\to
0$ yields the integral equation for the limiting effective field
\be\label{e45} u(\hat{x})=u_0(\hat{x})+i\xi
\int_Dg(\hat{x},\hat{y})N(\hat{y}) u(\hat{y})d\hat{y}. \ee One
obtains system \eqref{e40} if one solves equation \eqref{e45} by a
collocation method. Convergence of this method to the unique
solution of equation \eqref{e45} is proved in \cite{R563}. Existence
and uniqueness of the solution to equation \eqref{e45} are proved as
in \cite{R509}, where a three-dimensional analog of this equation
was studied.

One has $(\Delta_2+\kappa^2)g(\hat{x},\hat{y})=-\delta
(\hat{x}-\hat{y})$. Using this relation and applying the operator
$\Delta_2+\kappa^2$ to equation \eqref{e45} yields the following
differential equation for $u(\hat{x})$: \be\label{e46}
\Delta_2u(\hat{x})+\kappa^2u(\hat{x})+ i \xi N(\hat{x})u(\hat{x})=0
\quad \hat{x}\in \R^2. \ee This is a Schr\"{o}dinger-type equation,
and $u(\hat{x})$ is its scattering solution corresponding to the
incident wave $u_0=e^{i\kappa y}$.

Let us assume that $N(x)=N$ is a constant. One concludes from \eqref{e46}
that the limiting medium, obtained by
embedding many perfectly conducting circular cylinders, has
new parameter $\kappa_N^2:=\kappa^2+i\xi N$. This means
that $k^2=\kappa^2 +k_3^2$ is replaced by $\tilde{k}^2:=k^2+i\xi N$.
The quantity $k_3^2$ is not changed. One has $\tilde{k}^2=\omega^2n^2$,
$k^2=\omega^2 n_0^2$. Consequently, $n^2/n_0^2= (k^2+i\xi N)/k^2$.
Therefore, the  new
refraction coefficient $n^2$ is
\be\label{e47} n^2=n_0^2(1+i\xi N k^{-2}), \quad \xi=\frac{\omega \mu
\kappa^2}{\zeta k^2}.
\ee
Since the number $N>0$ and the impedance $\zeta$  are at our
disposal, equation \eqref{e47} shows that choosing suitable
$N$ one can create a medium with a desired  refraction
coefficient.

In practice  one does not go to the limit $a\to 0$, but
chooses a sufficiently small $a$. As a result, one obtains a medium
with a refraction coefficient $n^2_a$, which differs from
\eqref{e47} a little, $\lim_{a\to 0} n_a^2=n^2.$

\section{Conclusions}
Asymptotic, as $a\to 0$, solution is given of the EM wave scattering
problem by many perfectly conducting parallel cylinders of radius
$a$. The equation for the effective field in the limiting medium
obtained when  $a\to 0$ and the distribution of the embedded cylinders
is given by formula \eqref{e2}. The presented theory gives formula
\eqref{e47}
for the refraction coefficient in the limiting  medium. This formula shows
how the distribution of the cylinders
influences the refraction coefficient.

\section{Appendix}
Let us derive formulas \eqref{e14}-\eqref{e15}. Look for the
solution to \eqref{e3}-\eqref{e4} of the form: \be\label{eA1}
E_1=e^{ik_3z}\tilde{E}_1(x,y),\,\,\,
E_2=e^{ik_3z}\tilde{E}_2(x,y),\,\,\, E_3=e^{ik_3z}u(x,y), \ee
\be\label{eA2} H_1=e^{ik_3z}\tilde{H}_1(x,y),\,\,\,
H_2=e^{ik_3z}\tilde{H}_2(x,y),\,\,\, H_3=0, \ee where $k_3=const$.
Equation \eqref{e3} yields \be\label{eA3}
u_y-ik_3\tilde{E}_2=i\omega \mu \tilde{H}_1,\,\,\,
-u_x+ik_3\tilde{E}_1=i\omega \mu \tilde{H}_2,\,\,\,
\tilde{E}_{2,x}=\tilde{E}_{1,y}, \ee where, e.g.,
$\tilde{E}_{j,x}:=\frac{\partial \tilde{E}_{j}}{\partial x}$.
Equation \eqref{e4} yields \be\label{eA4} ik_3\tilde{H}_2=i\omega
\epsilon \tilde{E}_1,\,\,\, ik_3\tilde{H}_1=-i\omega \epsilon
\tilde{E}_2,\,\,\, \tilde{H}_{2,x}-\tilde{H}_{1,y}=-i\omega \epsilon
u. \ee Excluding $\tilde{H}_j$, $j=1,2$, from \eqref{eA3} and using
\eqref{eA4}, one gets \be\label{eA5}
\tilde{E}_1=\frac{ik_3}{\kappa^2}u_x,\,\,\,
\tilde{E}_2=\frac{ik_3}{\kappa^2}u_y,\,\,\, \tilde{E}_3=u, \ee
\be\label{eA6} \tilde{H}_1=\frac{k^2u_y}{i\omega
\mu \kappa^2},\,\,\, \tilde{H}_2=-\frac{k^2u_x}{i\omega
\mu \kappa^2}u_x,\,\,\, \tilde{H}_3=0. \ee Since
$E_j=\tilde{E}_j e^{ik_3z}$ and $H_j=\tilde{H}_j e^{ik_3z}$,
formulas \eqref{e14}-\eqref{e15} follow immediately from
\eqref{eA5}-\eqref{eA6}.

\end{document}